\documentclass[numreferences]{kluwer}

\usepackage{bm,epsfig}

\begin{document}
\begin{article}

\begin{opening}

\title{\center{
  Simple analytical model for entire turbulent }
  \\
  boundary layer over flat plane}
\subtitle{from viscous and mixing layers to turbulent logarithmic region}

\author{Victor S. \surname{L'vov}\email{victor.lvov@weizmann.ac.il}}
\author{Anna \surname{Pomyalov}\email{anna.pomyalov@weizmann.ac.il}}
\author{Vasil \surname{Tiberkevich}\email{vasyl.tyberkevych@weizmann.ac.il}}
\institute{
  Department of Chemical Physics,
  The Weizmann Institute of Science,
  Rehovot 76100, Israel
}

\runningtitle{Analytical model for TBL}
\runningauthor{L'vov et al.}

\begin{abstract}
  We discuss a simple analytical model of the turbulent boundary
  layer (TBL) over flat plane. The model offers an analytical
  description of the profiles of mean velocity and turbulent
  activity in the entire boundary region, from the viscous
  sub-layer, through the buffer layer further into the log-law
  turbulent region.  In contrast to various existing interpolation
  formulas the model allows one to generalize the description of
  simple TBL of a  Newtonian fluid for more complicated flows of
  turbulent suspensions laden with heavy particles, bubbles,
  long-chain polymers, to include the gravity acceleration, etc.
\end{abstract}


\abbreviations{
\abbrev{PBL}{planetary boundary layer};
  \abbrev{TBL}{turbulent boundary layer}; \abbrev{NSE}{Navier-Stokes
  equation};
  \abbrev{DNS}{direct numerical simulations};  \abbrev{LHS}{left-hand side};
    \abbrev{RHS}{right-hand side}; \abbrev{rms}{root mean square}
}


\end{opening}

Various problems of environmental and engineering hydrodynamics
call for a simple analytical model for the TBL over flat plane,
that can adequately describe from a unified viewpoint the mean
velocity and turbulent activity  profiles in the entire boundary
layer. In this paper we analyze in details such a model, announced
in \cite{drag1} in connection with a problem of drag reduction in
dilute polymeric solutions.  The model is based on the balance
equations for  mechanical momentum and kinetic energy. Aiming
maximum possible simplicity of the model we neglect the spacial
energy transfer in favor of the energy production and dissipation.
This makes our model local from a physical viewpoint and algebraic
from an analytical side. We also suggested a closure that links
the Reynolds stress with the density of kinetic energy. In
contrast to known interpolations between the {\sl resulting
formulas for the mean velocity profile} in the viscous and
turbulent sublayers we suggest a uniform model {\sl for the rate
of energy dissipation at the point of the formulation of the
model}. Besides the physical transparency, our approach allows
straightforward generalization of the model for more complicated
flows of turbulent suspensions laden with heavy particles,
bubbles, or long-chain polymers, inclusion of the gravity
acceleration, etc.

The basic Navier-Stokes equation (NSE) for the fluid velocity $\bm
U(\bm r,t)$  can be written as
\begin{equation}
\rho\left[ \frac{\partial \bm U}{\partial t} +\bm U\cdot \bm \nabla
\bm U\right] =-\bm \nabla p  +\mu \nabla^2 \bm U \ , \label{FP}
\end{equation}
where $\rho$ is the fluid (air) density, $p=p(\bm r,t)$ -- the pressure
and $\mu$ is the dynamical viscosity.  In this paper we follow the
standard strategy of Reynolds, considering  velocity as a sum of its
average (over time)  and a fluctuating part:
$$
\bm U(\bm r,t) = \bm{V}(\bm r)  + \bm u(\bm r,t) \ , \quad \bm
V(z) \equiv \langle \bm U(\bm r,t) \rangle \ . $$
 The objects that
enter our model in the planar geometry are the mean shear $S(z)$,
the Reynolds stress $W(z)$ and the kinetic energy $K(z)$; these
are defined respectively as
    \begin{equation}\label{object1}
 S(z)\equiv  \frac{d V_x(z)}{d z }\,,\quad
 W (z)\equiv - \rho \langle u_xu_z\rangle \,
, \ K(z) = \frac{\rho}{2}\langle |\bm u|^2\rangle\  .
    \end{equation}%
Here $x$, $y$, and $z$  are (horizontal) streamwise, spanwise, and
(vertical) wall-normal directions.

Integrating the stationary  NSE for the mean velocity $\bm{V}(z)$,
one gets a well known exact relation \cite{00Pope}, that describes
the point-wise balance of the flux of mechanical momentum:
\begin{equation}
\label{MF} \mu S(z) + W(z) = \mathcal{P}(z)\ .
\end{equation}
In the RHS of this equation we see the total flux of the mechanical
momentum $\mathcal{P}$; in the LHS we have the Reynolds stress and the
viscous contribution to the momentum flux. At large Reynolds numbers
one can usually neglect near the surface the production of
$\mathcal{P}(z)$ due to the pressure gradient or by some other
reasons. If so,
\begin{equation}\label{bound}
\mathcal{P}(z)=\mathcal{P}_0\equiv
\mathcal{P}(0)\ .
    \end{equation}
Having in mind  a  lower part of the planetary boundary layer
(PBL), we consider Eq.~(\ref{bound}) as a boundary condition at
the ground level $z=0$ instead of a given value of a free stream
velocity at the  upper boundary of PBL. The value of
$\mathcal{P}_0$  gives natural ``wall units" $ u_\tau, \tau$ and
$\ell_\tau$ for the velocity, the time and the length:
$$
    u_\tau\equiv\sqrt{\frac{\mathcal{P}_0}{\rho}}\,,
     \quad \tau\equiv \frac{\mu}
    {\mathcal{P}_0}
    \,, \quad \ell_\tau\equiv
    \frac{\mu}{\sqrt{ \rho\, \mathcal{P}_0}}\ .
 $$
Introducing so-called ``wall normalized" dimensionless objects
\begin{equation}\label{dim-l}
z^+\equiv\frac{z}{\ell_\tau}\,,\quad
V^+\equiv\frac{V_x}{u_\tau}\,, \quad
S^+\equiv\frac{S\,\ell_\tau}{u_\tau}\,, \quad W^+
\equiv\frac{W}{\rho\,u_\tau^2}\,, \quad
K^+\equiv\frac{K}{\rho\,u_\tau^2}\,,
\end{equation}%
we can rewrite Eq.~(\ref{MF}) as
\begin{equation}
\label{bal1} S^+(z^+) + W^+(z^+) = 1\ .
\end{equation}

A second relation between $S(z)$, $W(z)$ and $K(z)$ is
obtained from the ``point-wise" energy balance:
\begin{equation}\label{en-b}
\epsilon_{\rm prod}=\epsilon_{\rm dis}\,,\quad
\mbox{locality approximation.}
\end{equation}%
in which  we neglected the spacial energy transfer term,
$\epsilon_{\rm tr}$. The detailed analysis, see, e.g. Fig.~3 in
Ref.~\cite{DNS}, shows that in the log-law turbulent region this
term is small with respect to the energy dissipation term
$\epsilon_{\rm dis}$: $\epsilon_{\rm tr}\lesssim 0.1
\,\epsilon_{\rm dis}$. Clearly, in the viscous sub-layer the mean
velocity is  fully determined by the viscous term and thus the
influence of the energy transfer term can be neglected. For
simplicity of the model we will neglect $\epsilon_{\rm tr}$ term
also in the buffer layer (where the ratio $\epsilon_{\rm prod}\big
/ \epsilon_{\rm dis}$ is between 1 and 1.8).  We will show below
that the locality approximation~(\ref{en-b}) does not essentially
affect  the resulting mean velocity profile and Reynolds stress.

The energy production rate $ \epsilon_{\rm prod}$ in
Eq.~(\ref{en-b}) describes the energy flux from the mean shear
flow to the turbulent subsystem. In the plane geometry it has a
simple (and well known) form that follows from the NSE~(\ref{FP}):
\begin{equation}\label{en-pr}
\epsilon_{\rm prod}=W(z) S(z)\ .
\end{equation}%
It is also well known that the kinetic energy $K$ dissipates due
to viscosity at the rate
$$ 
\epsilon_{\rm dis}(z) = \mu \left \langle \left( \frac{ \partial
u_i}{\partial x_j}
 \right)^2 \right\rangle\ .
$$ 
In the viscous sub-layer  the velocity field is rather smooth, the
gradient exists and thus can be reasonably estimated via the
distance to the surface  as $1/z$. In other words, in this region
we can write
\begin{equation}\label{est1}
\epsilon_{\rm dis} \Rightarrow \epsilon_{\rm dis}^{\rm vis}(z)\,,
\end{equation}%
where
\begin{equation}\label{dis-vis}
\epsilon_{\rm dis}^{\rm vis}(z)
 \simeq \nu \left (\frac{a}{z}\right)^2 K(z)\,,\quad \nu\equiv\frac{\mu}
 {\rho}\,,
\end{equation}%
with  $a$ being  some dimensionless phenomenological constant of the
order of unity.

In the  buffer sublayer and in log-law turbulent region the energy
cascades down scales and is finally dissipated at the  Kolmogorov
(inner) scale that is much smaller than the distance $z$.
Therefore the contribution to the energy dissipations from all
scales, smaller  than $z$, is equal to the energy flux, which we
denote as $\epsilon_{\rm flux}$. This means that outside the
viscous sublayer
 Eq.~(\ref{est1}) has to be supplemented by an additional term,
$\epsilon_{\rm flux}$:
\begin{equation}\label{est2}
\epsilon_{\rm dis}(z) =  \epsilon_{\rm dis}^{\rm vis}(z)
+\epsilon_{\rm flux}(z) \ .
\end{equation}%
Notice, that in the buffer sublayer, both contributions in
(\ref{est2}) to $\epsilon_{\rm dis}(z)$ are equally important,
while in the log-law turbulent region the direct dissipation of
energy of turbulent eddies of the largest scale $z$ in the system,
given by Eq.~(\ref{dis-vis}), is negligibly small with respect to
the nonlinear energy flux $\epsilon_{\rm flux}$. Clearly, in the
viscous sublayer, Eq.~(\ref{est2}) also should work, because the
nonlinear contribution, $\epsilon_{\rm flux}(z)$ is negligibly
small with respect to the linear one, $ \epsilon_{\rm dis}^{\rm
vis}(z)$. We believe that Eq.~(\ref{est2}) is more than just
interpolation for the energy dissipation between the viscous
sublayer and log-law turbulent region. As we will show below, the
model~(\ref{est2}) gives an uniformly reasonable description of
the rate of the energy dissipation in the entire boundary layer.

To make this description constructive  one has to evaluate in
Eq.~(\ref{est2}) the energy flux  $\epsilon_{\rm flux}(z)$. This
can be done by standard Kolmogorov 1941-type  dimensional
reasoning:
\begin{equation}\label{est3}
\epsilon_{\rm flux}(z) \simeq \frac{K(z)}{\tau(z)} \simeq
\frac{b\, K(z)}{z}\sqrt{\frac{K(z)}{\rho}}\ .
\end{equation}%
Here $\tau(z)$ is the typical eddy turnover time at the height $z$
equal to the turnover time of the largest eddies (of scale~$z$) at
this height and $b$ is another dimensionless constant of the order
of unity. Thus Eqs.~(\ref{en-pr}, \ref{dis-vis}, \ref{est2}) and
(\ref{est3}) allows us to rewrite the energy balance
Eq.~(\ref{en-b}) in the entire boundary layer as follows:
 \begin{equation}\label{est4}
W(z)S(z)=\left[ \, \nu\, \left( \frac{a}{z} \right)^2+
\frac{b}{z}\sqrt{\frac{K(z)}{\rho}} \, \right]K(z)\ .
 \end{equation}%
In the dimensionless ``wall-normalized" objects~(\ref{dim-l}) this
equation reads:
 \begin{equation}\label{bal2}
W^+(z^+)S^+(z^+)=\left[ \, \left( \frac{a}{\ z^+} \right)^2+
\frac{b}{\ z^+}\sqrt{K^+(z^+)} \, \right]K^+(z^+)\ .
 \end{equation}%

Now we have two balance Eqs.~({\ref{bal1}}) and (\ref{bal2}) for
three objects, $S^+(z^+)$, $W^+(z^+)$ and $K^+(z^+)$. Two of them,
$W^+(z^+)$ and $K^+(z^+)$, are  different components of the same
Reynolds stress tensor $\langle u_i^+ u_j^+ \rangle$. Therefore it
is naturally to expect that in the scale invariant region (which
in our problem is the log-law turbulent region) these objects will
have the same $z$ dependence and thus their ratio will be
$z$-independent (dimensionless) constant:
\begin{equation}\label{frac}
\label{WK} \frac{W^+(z^+)}{K^+(z^+)} \equiv c^2(z^+)\Rightarrow
c^2_\infty\ .
\end{equation}
Notice that this ratio is bounded from above, $c^2(z^+)\le 1$,  by
the Cauchy-Schwarz inequality.  In fact the
expectation~(\ref{frac}) with $c_\infty^2\simeq 0.28$ in the
log-law turbulent region is in a good agreement with numerous
laboratory and nature experiments, see e.g. book~\cite{00Pope} and
with many DNS data, see for instance below Fig.~3,  taken from
Ref.~\cite{DNS}.

 Needless to say, that various Reynolds-stress based closure
procedures lead to the same result, $c_\infty=\ $const,  in the
log-law turbulent region, in which $c_\infty$ is expressed via yet
another phenomenological constants. In our simple model we prefer
to use Eq.~(\ref{WK}) as a basic closure. Moreover, we argue  that
we can safely use Eq.~(\ref{WK}) not only in the log-law turbulent
region, where it is definitely valid, but also in the buffer layer
and even in the viscous sublayer, where Eq.~(\ref{WK}) is
violated. The reason is simple: the larger the deviation of the
ratio ${W^+(z^+)}/ {K^+(z^+)}$ from the constant $c^2_\infty$, the
less important become relation~(\ref{WK}) itself in the momentum
and energy balances. Below in this paper we account  numerically
for the real $z^+$-dependence of the ratio ${W^+(z^+)}/
{K^+(z^+)}$ and demonstrate that this is insignificant for the
mean velocity and the Reynolds stress profiles.

Equation~(\ref{WK}) allows us to represent our model,
Eqs.~({\ref{bal1}}) and (\ref{bal2}), in terms of just two
unknowns, $S^+$ and $K^+$ or $W^+$. We choose  $W^+$ instead of
$K^+$, because the Reynolds stress is responsible for the
turbulent transport of the mechanical momentum and thus plays a
more important role in the wall bounded turbulence than the
kinetic energy. Notice, that in the wall turbulence $W^+$ is
positive definite, since the momentum flux is directed toward the
surface.

In terms of $S^+$ and $W^+$ Eqs.~({\ref{bal1}}) and  (\ref{bal2})
read:
\begin{eqnarray}\label{mod1}
1&=&S^+(z^+)+ W ^+(z^+)\,,\\ \label{mod2}
0&=&\left[ c_\infty^2 S^+(z^+) -\Gamma^+(z^+)\right]W^+ (z^+)\ .
\end{eqnarray}%
Here $\Gamma^+(z^+)$ can be considered as an  effective damping
rate of the turbulent fluctuations:
\begin{equation}\label{G}
\Gamma^+(z^+)\equiv
\left( \frac{a}{\ z^+} \right)^2+
\frac{b}{c_\infty\, z^+}\sqrt{W^+(z^+)}\,,
\end{equation}%
and $ c_\infty^2 S^+(z^+)$ clearly represent the energy influx
rate.

The basic equations of our model (\ref{mod1}) and (\ref{mod2})
have two solutions: a laminar and a turbulent one. In the laminar
solution there are no turbulent fluctuations:
\begin{eqnarray}\label{lam}
 W^+(z^+)&=&K^+(z^+)= 0\,,
\\ \nonumber  S^+(z^+)&=&1\,,\qquad  V^+(z^+)=z^+\,:
 \quad \mbox{laminar solution.}
\end{eqnarray}%
The stability condition with respect to appearance of the
turbulent fluctuations requires that the damping rate,
$\Gamma^+(z^+)$,  at a zero level of turbulence is larger (or
equal) than the pumping rate, $ c_\infty^2 $ [recall, that in the
laminar solution  $ S^+(z^+)=1 $]:
\begin{equation}\label{stab}
\left( \frac{a}{z^+}\right)^2 \ge c_\infty^2
 \,, \qquad \mbox{stability of the laminar solution.}
\end{equation}%
This equation shows that the laminar solution (\ref{lam}) is stable
near the surface, for $z^+\le z^+_{\rm vs}$, where in our model
\begin{equation}\label{stab1}
z^+_{\rm vs}\equiv \frac{\, a}{c_\infty}
 \quad \mbox{is the upper  boundary of the
viscous-sublayer.}
\end{equation}%
Recall, that  the
energy transfer is neglected in the model. Therefore it is not
surprising that Eq.~(\ref{lam}) demonstrate no turbulent activity
in this sublayer.

There is however a turbulent activity in the rest of the boundary layer:
\begin{equation}\label{turb1}
~~~~~~~~W^+(z^+)>0\,, \ \mbox{for}\ z^+>z^+_{\rm vs}\,,\
\mbox{mixing layer \& log-law region,}
\end{equation}%
in which  Eq.~(\ref{mod2}) gives
 $$c_\infty^2 S^+(z^+)= \Gamma^+(z^+)\ .
 $$
This relation together with definition (\ref{G}) yield:
$$
c_\infty ^2\,S^+(z^+) =\left(
\frac{a}{\ z^+} \right)^2+ \frac{b}{c_\infty\,
z^+}\sqrt{W^+(z^+)}\ .
$$ 
Dividing this equation  by $c^2_\infty$, and using definition
(\ref{stab1}) for $z^+_{\rm vs}$, one finally gets instead of
(\ref{mod1}, \ref{mod2}) a new set of coupled equations:
\begin{eqnarray}\label{turb3}
1&=&S^+(z^+)+ W ^+(z^+)\,,\\
S^+(z^+) &=&\left( \frac{z^+_{\rm vs} }{\ z^+} \right)^2+
\frac{\sqrt{W^+(z^+)}}{\kappa \, z^+}\ .
\label{turb4}
\end{eqnarray}%
Here we introduced another dimensionless parameter  $\kappa$
\begin{equation}\label{kappa}
\kappa\equiv\frac{c_\infty^3}{b}\,,
\end{equation}%
which in our model is nothing but the von-Karman constant, that
defines the slope of the logarithmic mean velocity profile in the
log-law turbulent region. Notice, that the final system of coupled
equations (\ref{turb3}) and (\ref{turb4}) have a minimum possible
number (just two) of phenomenological constants, $z^+_{\rm vs}$
and $\kappa$ (that are some combinations of initially introduced
three parameters, $a$, $b$ and $c_\infty$). Indeed, any models of
wall bounded turbulence have at least two  phenomenological
parameters, see,e.g. \cite{00Pope}.  For example, the famous
``logarithmic law of the wall"
\begin{eqnarray}\label{K-prof}
 V^+(z^+) &=&\kappa^{-1}\ln z^+ + B\,,
 \quad{\rm for}~ z^+ \gtrsim 30 \,,\\ \nonumber
\kappa&\approx& 0.436\,, \quad B\approx 6.13\,,
\end{eqnarray}%
contains  the von-Karman constant $\kappa$ and the intercept $B$,
with experimental values in Eq.~(\ref{K-prof}) taken from
\cite{97ZS}.

Let us show, that unlike (\ref{K-prof}), Eqs.~(\ref{turb3},
\ref{turb4}) describe the velocity profile in the entire boundary
layer and not only in the log-law turbulent region. Eliminating
$W^+$ from Eqs.~(\ref{turb3}, \ref{turb4}) one gets a quadratic
equation for $S^+$ with two solutions. The physical one has the
form:
\begin{equation}\label{sol1}
  S^+ (z^+)= \frac{2\kappa^2\left(z_{\rm vs}^+\right)^2-1
               +\sqrt{4\kappa^2
               \left[{z^+}^2 -\left(z_{\rm vs}^+\right)^2\right]+1}}
        {2\kappa^2{z^+}^2} \ .\end{equation} Now Eq.~(\ref{turb3})
immediately gives an expression for $W^+$, which is valid for $ z^+
\ge z_{\rm vs}^+:$ 
\begin{equation}\label{sol2}
  W^+ = \frac{2\kappa^2\left[{z^+}^2 -\left(z_{\rm vs}^+\right)^2\right] +1
               -\sqrt{4\kappa^2\left[{z^+}^2 - \left(z_{\rm
vs}^+\right)^2\right]+1}} {2\, \kappa^2{z^+}^2} \ .
\end{equation}
 One sees that at $z^+=z_{\rm vs}^+$ the turbulent solution~(\ref{sol1},
\ref{sol2}) coincides with the laminar  solution (\ref{lam}):
$S^+(z_{\rm vs}^+)=1$, $W^+ (z_{\rm vs}^+)=0$, as expected. To get
the mean velocity profile, we integrate Eq.~(\ref{sol1}) matching
the result with the laminar solution~(\ref{lam}) at $z^+=z^+_{\rm
vs}$. Fortunately, the expression for the mean shear (\ref{sol1})
allows analytical integration. In is convenient to present the
result of this integration in the form, similar to the logarithmic
law of the wall (\ref{K-prof}):
\begin{equation}\label{V+}
V^+(z^+)=\kappa^{-1}\ln Z(z^+)+B - \Delta(z^+)\,,\qquad \mbox{for}\ z^+
\ge z_{\rm vs}^+\ .
\end{equation}%
Here  functions $Z(z^+)$, $\Delta(z^+)$ and intercept $B$ are
given by
\begin{eqnarray*}\label{Z}
&&\hskip -1cm Z(z^+)=\frac12\, \left[z^++\sqrt{{z^+}^2-{z_{\rm
vs}^+}^2 +\big(2\kappa\big)^{-2}}\ \right]\to z^+\quad \mbox{for}\
z^+\to \infty \,,\\ &&\hskip -1cm\label{D} \Delta (z^+)=\frac{2\,
\kappa^2 {z_{\rm vs}^+}^2+ 4 \kappa \big[Z(z^+)-z^+\big]-1}{2
\kappa^2 z^+}\quad \to 0\quad\ \mbox{for}\ z^+\to \infty\,,\\
&&\hskip -1cm \Big. B= 2z^+_{\rm vs}-\kappa^{-1} \ln\left[{e\, (1+
2\,\kappa z^+_{\rm vs})}/{4\kappa}\right]\ .\label{B}
\end{eqnarray*}

Note that Eq.~(\ref{V+}) pertain to the whole $z^+\ge z^+_{\rm
vs}$ domain, meaning both mixing sublayer and log-law turbulent
region. By taking the experimental values of $\kappa$ and $B$ we
compute $z^+_{\rm vs}\approx 6$ to be compared with the
experimental value of $5.5\pm 0.5$, cf. \cite{00Pope}. The
resulting mean velocity profile for the entire boundary layer,
Eqs.~(\ref{lam}) and (\ref{V+}), is shown in Fig.~1 as the solid
line. The excellent agreement with the experimental and numerical
data in the entire region of $z^+$ indicates that our balance
equations are sufficiently accurate.

\begin{figure}\label{f:prof1}
  \centering\includegraphics[width=1.1 \textwidth]{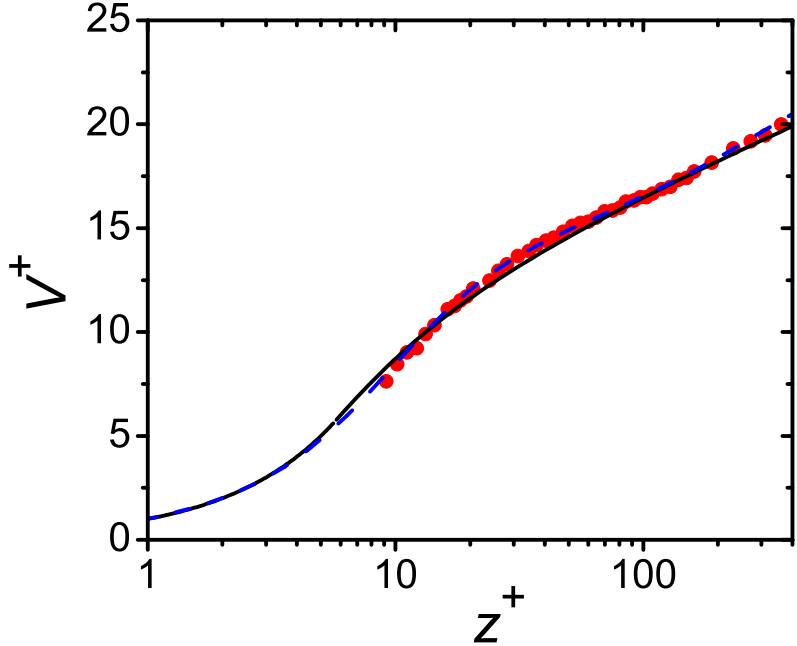}
  \caption{
    Mean velocity profiles $V^+(z^+)$: black  solid line -- our
    analytical model, Eqs.~(\ref{lam}) and (\ref{V+}); blue dashed
    line -- results of the DNS simulation \cite{DNS}; red points
    -- experimental data \cite{97ZS}.}
\end{figure}

In Fig.~2 we show analytical profiles of the Reynolds stress
$W^+(z^+)$, Eq.~(\ref{sol2}), in comparison with DNS. Due to the
limited value of the friction Reynolds number in the DNS
data~\cite{DNS}, Re$_\lambda=590$,  we normalized the Reynolds
stress  using \emph{the local value of the momentum flux}
$$ 
 \mathcal{P}(z^+)=\mathcal{P}_0\Big(
1-\frac{z^+}{\mbox{Re}_\lambda} \Big)\ .
$$ 
The same normalization was used in Fig.~5 for the kinetic energy.
Notice that the type of normalization [with $\mathcal{P}_0$ or
$\mathcal{P}(z^+)$] does not affect  the mean profile (due to its
slow, logarithmic dependence on $z^+$). Therefore, in the DNS data
for mean velocity in Fig.~1 we used the simple normalization with
$\mathcal{P}_0$. One can see in Fig.~2 an excellent agreement of
our analytical results with the  DNS results. The minor
discrepancy is observed only in the viscous sublayer, $z^+\le
z^+_{\rm vs}$: in our simple approach the Reynolds stress and the
turbulent kinetic energy are identically zero in this region, see
Eq.~(\ref{lam}). This stems from the disregard of the energy
transfer term in the energy balance equation (\ref{en-b}), which
gives a non-zero level of turbulent activity close to the surface.
As one sees however from Figs.~1 and~2,  the account of the
spacial transfer term does not lead to a considerable value of the
Reynolds stress in the viscous sublayer and does not affect the
mean velocity at all.

\begin{figure}\label{f:prof2}
  \centering\includegraphics[width=\textwidth]{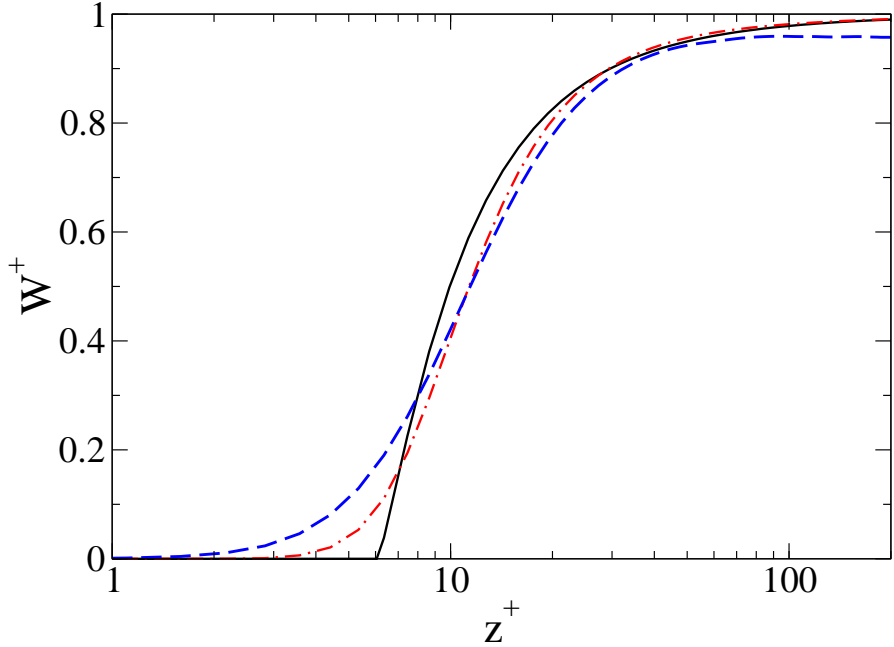}
  \caption{
    Profiles of the Reynolds stress $W^+(z^+)$: black solid line
    -- our analytical model, Eqs.~(\ref{sol2}) with $c^2_\infty=0.28$,
    red dash-dotted line --  Eqs.~(\ref{sol2})
    with function $c(z^+)$, given by Eq.~(\ref{fit}),  and  blue
    dashed line -- results of the DNS simulation \cite{DNS}.}
\end{figure}

Another assumption that fails in the viscous and buffer layers is the
approximation of constancy of the correlation coefficient
$c^2_\infty\equiv W^+/K^+$. Note, however, that the expressions for
the mean shear (\ref{sol1}) and the Reynolds stress (\ref{sol2})
remain valid even for $z^+$-dependent correlation coefficient
$c_\infty\Rightarrow c(z^+)$. In this case $z^+_{\rm vs}$ and $\kappa$
should be understood as $z^+$-dependent functions:
$$
  z^+_{\rm vs} \Rightarrow z^+_{\rm vs}(z^+) \equiv \frac{a}{c(z^+)}
\,,\qquad
  \kappa \Rightarrow \kappa(z^+) \equiv \frac{c^3(z^+)}{b}
\ .$$
 In fact, we have chosen $c(z^+)={\rm const}$ only to  make
possible an analytic expression for the mean velocity profile,
Eq.~(\ref{V+}). In our model we can easily account for the
``realistic" $z^+$-dependence of the ratio $c^2$  integrating
Eq.~(\ref{sol1}) numerically.

The actual dependence $c(z^+)$, shown in Fig.~3 by blue dashed
line,  is taken from the public available statistical database,
produced in Ref.~\cite{DNS} by  DNS of the NSE for high-Reynolds
turbulent flow in the channel geometry. One sees that $c(z^+)$
decreases toward the surface. This fact can be understood by a
series expansion for $W^+(z^+)$ and $K^+(z^+)$ for $z^+\to0$ (see,
e.g. \cite{00Pope}).  This expansion shows that near the surface
$W^+(z^+)$ and $K^+(z^+)$ behave as $$
  W^+(z^+) \sim ({z^+})^3
\,,\qquad
  K^+(z^+) \sim ({z^+})^2 \,,$$ 
  and therefore
  \begin{equation}\label{as1}
c^2(z^+)\sim z^+ \,, \quad \mbox{near the surface.}
  \end{equation}%

An origin of these dependencies is quite simple. At  the surface,
the rms values of the horizontal projections of the turbulent
velocity $\sqrt{\langle {u_x^+}^2 \rangle}$ and $
\sqrt{\langle{u_y^+}^2 \rangle}$  are zero according to the
no-slip boundary conditions and  grow with the height as $z^+$.
This is not the case for the  vertical projection,
$\sqrt{\langle{u_z^+}^2 \rangle}$. The  incompressibility
constraint dictates that
$$\frac{\partial u_z}{\partial z}
= - \left( \frac{\partial u_x}{\partial x}+ \frac{\partial
u_y}{\partial y}\right) \propto z$$ and therefore the vertical
projection increases only as $z^2$:
$$
\sqrt{\left\langle {u_x^+}^2 \right\rangle}\sim
\sqrt{\langle{u_y^+}^2 \rangle} \sim z^+\,,\quad
\sqrt{\left\langle {u_z^+}^2 \right\rangle}\sim ({z^+})^2\ .
$$
Accordingly,
$$
W^+(z^+)\equiv-\left\langle u^+_x\, u^+_z \right\rangle\sim
\sqrt{\langle {u_x^+}^2\rangle\,\langle {u_z^+}^2\rangle}\sim
({z^+}) ^3\,,
$$
while the kinetic energy is dominated by the horizontal turbulent
velocities,
$$
  K^+(z^+)\simeq \frac12\left[ \langle {u_x^+}^2\rangle+
  \langle {u_y^+}^2\rangle  \right]\sim  ({z^+}) ^2\,,
$$
in agreement with Eq.~(\ref{as1}).

To demonstrate how the actual dependence $c^2(z^+)$ influence the
mean velocity and turbulent activity profiles we suggest the
following interpolation formula,
\begin{equation}\label{fit}
c^2(z^+)=c^2_\infty\left[ 1-\exp \Big( - \frac{\ z^+}{z^+_{\rm
cr}} \Big) \right]\,,
\end{equation}%
that has just two parameters, asymptotic value $c_\infty^2$ for
$z^+\to \infty$ and the slope $c_\infty^2/z^+_{\rm cr}$ of the
linear dependence~(\ref{as1}) for $z^+\to 0$, near the surface.
Dependence~(\ref{fit}) with $c_\infty^2=0.28$ and $z^+_{\rm
cr}=24$ is shown in Fig.~3 by red dash-dotted line. One sees that
Eq.~(\ref{fit})  closely fits the DNS data in the entire boundary
layer and thus can be used as a realistic representation of the
$W^+/K^+$ ratio in our model to find  the analytical
representation for the improved profiles of the Reynolds stress
and kinetic energy, Eqs.~(\ref{frac}, \ref{sol2}) and then,  after
numerical integration of Eq.~(\ref{sol1}) to get  the improved
mean velocity profile.

\begin{figure}
\centering \includegraphics[width= 0.9 \textwidth]{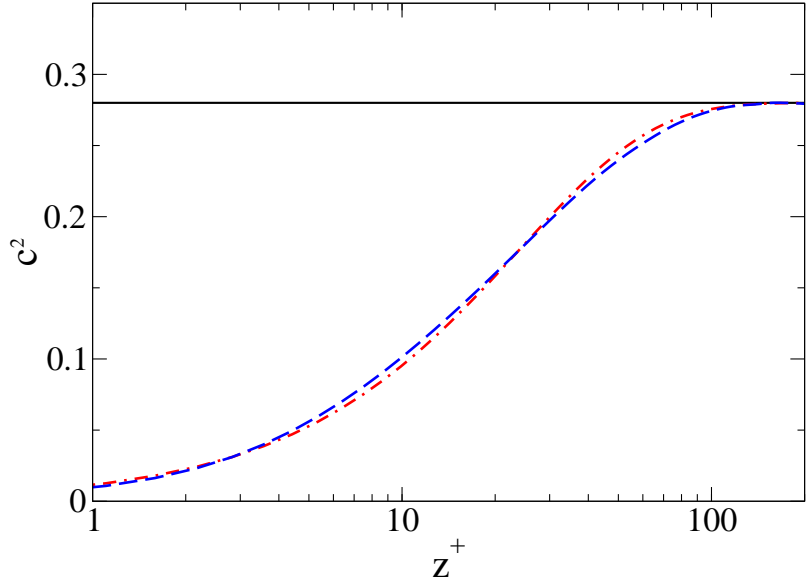}
\caption{Blue dashed line -- DNS data~\cite{DNS} for $c^2(z^+)$,
red dash-dotted line -- suggested fit Eq.~(\ref{fit}) for
$c^2(z^+)$,  and horizontal black solid line -- asymptotical value
$c^2_\infty=0.28~\cite{DNS}$.}
\end{figure}
\begin{figure}
\includegraphics[width= 0.9 \textwidth]{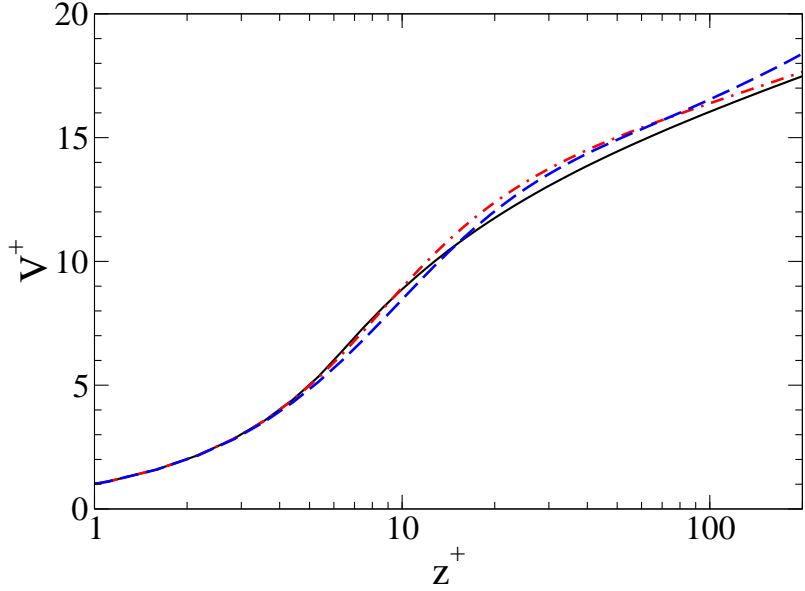}
\caption{Mean velocity  profiles
$V^+(z^+)$:  black solid line --  analytical profile, Eq.~(\ref{V+})
 with $a=3.2$, $b=0.27$ and $c=c_\infty$ (the same, as in
Figs.~1-3) and  red dash-dotted line -- result of the numerical
integration of Eq.~(\ref{sol1}) with  fit function~(\ref{fit}) for
$c(z^+)$ and $a=0.3$ and the same value of $b=0.27$. Blue dashed
line -- DNS data~\cite{DNS}.}
\end{figure}

The comparison of the resulting profiles is given in Fig.~2
(Reynolds stress), Fig. 4 (mean velocity) and Fig 5 (kinetic
energy). The profiles in the simple model with $c(z^+)\Rightarrow
c_\infty$ are denoted by black solid lines, the improved profiles
with $z^+$-dependent coefficient $c(z^+)$ -- by red dash-dotted
lines and the DNS profiles -- by blue dashed lines. One sees in
Fig.~4 that all mean velocity profiles nearly collapse: our
approximation has no effect on the function $V^+(z^+)$. For this
object we prefer to take $c(z^+)=c_\infty$ and to have a fully
analytic model.  The profile of the Reynolds stress, as one sees
in Fig.~2, is improved in the viscous sublayer and is affected
very little in the rest of boundary layer by account for
$z^+$-dependence of $c(z^+)$. As for the profile of the kinetic
energy, Fig. 5, the neglect of the actual $z^+$-dependence of the
coefficient $c(z^+)$ leads to a significant underestimate of the
kinetic energy in the buffer sublayer. In particular, the simple
model does not exhibit a peak of $K^+(z^+)$ in this region. If
this peak is essential for some particular reasons, one should
account for the actual $z^+$-dependence of $c(z^+)$ at the expense
of simplicity.

We have to stress, that the effect of the  spacial energy
transfer, neglected in our simple approach, is absolutely
insignificant for the mean velocity profile (see Figs. 1 and 4).
It has a minor importance for the profile of the Reynolds stress
(Fig. 2) in the buffer layer and plays a more important role for
the profile of kinetic energy, decreasing the amplitude of  its
peak and increasing the value of $K^+(z^+)$ from the left of its
maximum, in the viscous sublayer.  Therefore, the necessity to
account for the transfer term should be evaluated for each
particular problem in hand.

\textbf{In conclusion}: In this paper we  discussed in details a
simple model of a turbulent flow  over a flat plane that offer an
analytical description of the mean velocity, Reynolds stress and
kinetic energy profiles in the entire boundary layer. The
calculated profiles exhibit an excellent agreement with the
results of the laboratory experiments~\cite{97ZS} and
DNS~\cite{DNS} of the NSE. We discussed the effect of the
approximations made in the analytical description of the profiles.
We found a simple functional form for the experimental
$z^+$-dependence of the correlation coefficient $c(z^+)$, that
allows to relax the approximation of the constancy of this
coefficient. The profiles calculated with the help of this
$z^+$-dependent coefficient $c(z^+)$ account for all physically
important features of all three profiles in the entire boundary
layer region. The physical transparency and simplicity of the
model allow its generalization for turbulently flowing
suspensions, as it was demonstrated on the example of the problem
of drag reduction by polymers in Refs.~\cite{drag1,drag2,drag3}.

\begin{figure}\label{f:control}
\centering \includegraphics[width=0.8\textwidth]{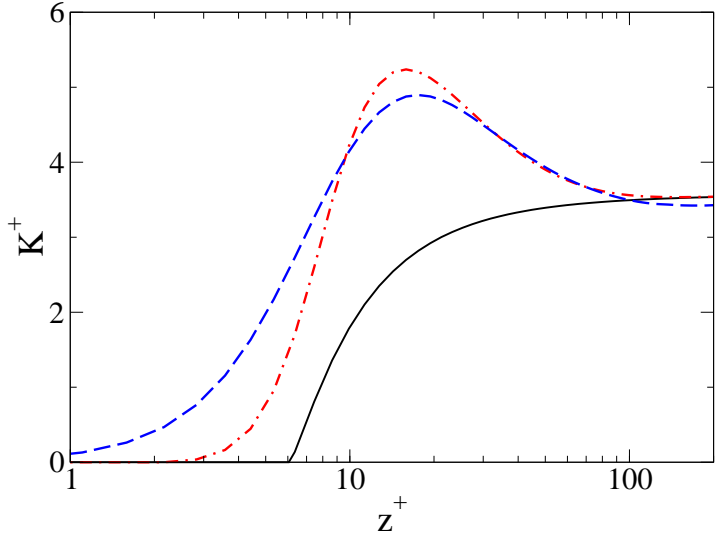}
\caption{Mean profiles of the kinetic energy $K^+(z^+)$:
  black solid line --  analytical profile with $c=c_\infty$
red dash-dotted line accounts for the $z^+$-dependence~(\ref{fit})
of $c$ and blue line -- is the DNS profile~\cite{DNS}.}
\end{figure}

\acknowledgements%
 We acknowledge useful discussions with Itamar Procaccia and
Sergej Zilitinkevich.  Our special thanks to R.~G.~Moser, J.~Kim,
and N.~N.~Mansour, who made their comprehensive DNS data of high
Re channel flow
public available in Ref.~\cite{DNS}. This work was supported by
the US-Israel Binational Scientific Foundation.

\end{article}
\end{document}